# Near-field imaging in the megahertz range by strongly coupled magnetoinductive surfaces: theoretical model and experimental validation

**Manuel J. Freire and Ricardo Marqués**
Departamento de Electrónica y Electromagnetismo,
Facultad de Física, Universidad de Sevilla, Avda. Reina Mercedes
s/n, E 41012 Sevilla, Spain

**Abstract**

In this work, near-field imaging by two strongly coupled arrays of split ring resonators is analyzed. A simple theoretical model is developed to obtain the transfer function of the lens. This model shows that magnetoinductive surface waves (MISWs) play the same role as plasmon-polaritons in negative refractive slabs. In particular, the model predicts that the image is formed near the resonant frequency of the rings, between the pass-bands for the two MISW branches that can be excited in the lens. It also predicts a significant image enhancement when the distance between the source source and the image plane is smaller than twice the lens width. The predictions of the theoretical model are supported by measurements in the radio-frequency range. This suggests the possibility of using this kind of devices to imaging processes in the megahertz range, as for example in magnetic resonance imaging.

## I. INTRODUCTION

After the seminal works of Veselago [1] and Pendry [2], imaging by metamaterial slabs became an active area of research. Although losses may substantially degrade this effect [3], they do not prevent the onset of sub-diffraction images in near field experiments [4-6]. In fact, near field sub-diffraction imaging by metamaterial slabs [2] has been reported in experiments carried out with a two dimensional circuit analogous of a left-handed medium [7], a single-cell-depth left-handed slab [8], a silver slab operating at optical frequencies [9], a pair of magneto-inductive surfaces [10], and a photonic crystal flat lens [11]. In addition, it has been theoretically predicted for ferrite slabs [12] and in a pair of coupled resonant impedance surfaces [13]. The main aim of this paper is to further develop the analysis of the magneto-inductive (MI) lens previously reported by the authors [10]. In particular, the role of MI surface waves for the image formation will be clarified, and the field distribution around the image will be studied. Some interesting physical effects, such as the image enhancement at the back side of the lens, are also reported.

The aforementioned MI lens is a near-field imaging device that can be designed to work at frequencies ranging from megahertz to terahertz. The MI lens consists of two parallel planar periodic arrays of inductively coupled resonators. A key concept in near-field sub-diffraction imaging is the amplification, inside the lens, of evanescent spatial Fourier harmonics (FHs) coming from the source. This amplification produce the restoration of the FHs amplitude at the image plane [2], thus allowing for the image formation. In the analyzed MI lens, such amplification process is closely related to the excitation of MI surface waves (MISWs) [14] at each individual array of resonators [10]. However, such MISWs do not correspond to the MISWs that can be excited in the device. In fact, when two arrays are coupled in order to form the MI lens, the MISWs dispersion relation splits into two branches, with pass-bands above and below the pass-band of the original MISW of the isolated array. The excitation of anyone of these MISW branches is undesirable for image formation, because it would imply radiation in the plane of the device, as well as image distortion by the disproportionate contribution of such resonances. A similar conclusion was previously reported regarding plasmon-polariton excitation in imaging devices made of negative permittivity slabs [15]. In order to avoid such undesirable effect, the coupling between both arrays of resonators must be strong enough to produce a clear separation between the pass-bands of both MISW branches. When this condition is achieved, the image is expected to appear at some frequency between the pass-bands for both MISW branches. For this reason, strongly coupled arrays of resonators are considered in this work.

In order to develop the analysis, a surface impedance model is proposed for each array of resonators. The transfer function of the lens is obtained following the method previously proposed by Maslovski[13]. The aforementioned MISW branches will appear as poles of this transfer function. Between such poles, a flat transfer function with an almost constant value near unity at the image plane will be found, thus justifying the image formation. In order to obtain the best resolution, the source plane will be located at a distance from the lens equal to the distance between the arrays. Thus, the image is formed just at the exit of the lens. This configuration reduces losses in the arrays, and is optimal from the resolution standpoint[16]. It has been previously used in other sub-diffraction image experiments, carried out in the optical range of frequencies [9]. The effect on the image of variations in the distance between the source and the lens is also studied. A significant image enhancement will be obtained when the source approaches the lens, in agreement with some previous theoretical results [17].

Experiments have been carried out in order to validate our theoretical analysis. Such experiments are aimed to the direct measurement of the field at the image plane, i.e. at the exit o the lens. Regarding such measurements, care must be taken in order to avoid artifacts coming from the matching capabilities of the lens [18,19]. It was already mentioned that sub-diffraction image formation is based on the amplification, inside the lens, of evanescent FHs coming from the source. Since there is no power flux associated with an evanescent FH, this amplification does not violates energy conservation [2]. However, any field measurement implies some flux of energy. Thus, if unappropriate detectors are used, the measurement procedure will substantially affect the field at the image, and the behavior of the lens may be closer to a tunneling device than to a imaging device[19]. Although such tunneling effect may be of interest for some applications, it must be avoided for the particular purpose of this work. I order to avoid such effect, it has been used small detectors with an impedance sufficiently smaller than the characteristic impedance of the input/output waveguides of the measurement device. This choice ensures that the measurements actually correspond to a direct measurement of the field created behind the lens by the source alone [19].

## II. THEORY

The MI lens analyzed in this work is shown in Fig. 1. It is formed by two planar periodic arrays of capacitively loaded open rings (CLORs), placed at a distance $d$ between them. The transfer function for the device is obtained following the procedure already reported by Maslovski [13]. Although the main steps for this computation were already described in such paper, they will be briefly outlined here for completeness. Using the same notation as in Maslovski's work [13], the complex amplitudes of waves travelling (or decaying) in opposite directions across the device are related by a 2x2 transmission matrix as:

$$\begin{bmatrix} E_2^- \\ E_2^+ \end{bmatrix} = T_{tot} \cdot \begin{bmatrix} E_1^- \\ E_1^+ \end{bmatrix} \quad T_{tot} = \begin{bmatrix} t_{11} & t_{12} \\ t_{21} & t_{22} \end{bmatrix} \tag{1}$$

where $E_1^{\pm}$ and $E_2^{\pm}$ denote the tangential components of the electric field complex mplitude at both the source and the image plane (subscript 1 denotes the source plane and subscript 2 denotes the image plane). The signs $\pm$ correspond to the signs in the exponential term $exp(\pm i k z)$ of these waves, $z$ being the axis perpendicular to the lens, and $k$ the propagation constant along this direction. The transfer function $T$ can be obtained form the above transmission matrix $T_{tot}$ as:

$$T = \frac{E_2^-}{E_1^-}\bigg|_{E_2^+=0} \tag{2}$$

In the experiments reported in this work, the image plane is fixed at the exit of the lens, and the source plane is placed at a certain distance from the lens (as it was already mentioned, this configuration is optimal from the resolution standpoint[16]). Taking this into account, the transmission matrix $T_{tot}$ can be written as

$$T_{tot} = T_{lens} \cdot T_{sp\,before} \tag{3}$$

Where $T_{lens}$ is the transmission matrix of the lens and $T_{sp\,before}$ represents the transmission matrix for the air layer occupying the space between the source plane and the lens. Denoting by $d$ the thickness of the lens, if the source plane is placed at the same distance, $d$, from the lens, $T_{sp\,before}$ is as follows

$$T_{sp\,before} = \begin{bmatrix} \exp(-jkd) & 0 \\ 0 & \exp(+jkd) \end{bmatrix} \tag{4}$$

where $k = \sqrt{k_0^2 - k_x^2 - k_y^2}$. Moreover, $T_{lens}$ can be written as

$$T_{lens} = T_{out} T_{sp(d)} T_{in} \tag{5}$$

where $T_{in}$ and $T_{out}$ are the matrices describing the two arrays of resonators. In our case both matrices are identical and can be written as [13]:

$$T_{in} = T_{out} = \begin{bmatrix} 1 - \dfrac{\eta_0}{2Z_g} & -\dfrac{\eta_0}{2Z_g} \\ \dfrac{\eta_0}{2Z_g} & 1 + \dfrac{\eta_0}{2Z_g} \end{bmatrix} \quad (6)$$

where $Z_g$ is the surface impedance of the array of resonators, and $\eta_0$ is the wave impedance of vacuum. The grid impedance $Z_g$ is expressed in function of the so called cell impedance $Zc$ as[13]

$$Z_g = Z_c - \frac{\eta_0}{2} \quad (7)$$

where $Zc$ is the impedance relating the average surface current density **J** on the array and the external electric field **E**$_{ext}$ imposed by the source, that is

$$\mathbf{E} = Z_c \mathbf{J} \quad (8)$$

In the MI lens the resonators are excited by the ]z-]component of the external magnetic field **H**$_{ext}$ imposed by the source. For our analysis it is then convenient to obtain the relation between $H_{ext,z}$ and **E**$_{ext}$. This relation directly follows from Faraday's law, i.e.:

$$H_{ext,x} = \frac{\mathbf{k}_t \times \mathbf{E}_{ext}}{\omega \mu_0} \cdot \mathbf{z} \quad (9)$$

where $\mathbf{k}_t = k_x \mathbf{x} + k_y \mathbf{y}$ is the wavevector along the array. Next, on the array, the averaged surface current density **J** can be obtained from the averaged magnetic moment **M** through **rot**[**M**] = **J**. Therefore

$$\mathbf{J} = -\frac{1}{a^2}(i\mathbf{k}_t \times \mathbf{m}) \quad (10)$$

where $a$ is the periodicity of the array and **m** is the magnetic moment of each resonator. Taking into account (8), (9) and (10), $Zc$ can be expressed as

$$Z_c = -\frac{i\omega \mu_0 a^2 H_{ext}}{m(k_x^2 + k_y^2)} \quad (11)$$

The magnetic moment and the external magnetic field are then related through

$$m = \chi(\omega)[H_{ext} + \beta(\omega, k_t)m] \quad (12)$$

where $\chi(\omega)$ is the resonator polarizability and $\beta(\omega,k_t)$ an interaction factor, which takes into account the inter resonators coupling. Thus, taking into account (11) and (12), $Zc$ can be written as

$$Z_c = -\frac{i\omega \mu_0 a^2}{k_x^2 + k_y^2}\left(\frac{1}{\chi(\omega)} - \beta(\omega, k_t)\right) \quad (13)$$

Next, it is shown that both χ(ω) and β(ω,$k_t$) can be derived from the dispersion relation of the MI waves propagating in a single 2D-array of resonators under the influence of the external magnetic field. This dispersion relation can be obtained from the circuit equation for a single resonator[14]

$$\left[R + i\omega L + \frac{1}{i\omega C}\right]I = i\omega M\, 2I\left[\cos(k_x a) + \cos(k_y a)\right] + i\omega\mu_0 H_{ext} S_0 \qquad (14)$$

where $I$ is the intensity of the ring, $R$, $L$ and $C$ are the resistance, self-inductance and capacitance of each resonator, respectively, and $M$ is the mutual inductance between neares resonators in the array. The first term on the right side of (14) accounts for the voltage induced on each resonator by the four nearest neighbors[14] whereas the second term accounts for the voltage induced by the external magnetic field.

Taking into account that m=$I\, S_o$, where $S_o$ is the effective ring surface, the magnetic moment can be obtained as follows:

$$m = \frac{\mu_0 S_0^2 \omega^2}{L(\omega_0^2 - \omega^2) + i\omega R}\left[H_{ext} + m\frac{2M}{\mu_0 S_0^2}(\cos(k_x a) + \cos(k_y a))\right] \qquad (15)$$

Comparing both (12) and (15) the following expressions for χ(ω) and β(ω,$k_t$) are obtained

$$\chi(\omega) = \frac{\mu_0 S_0^2 \omega^2}{L(\omega_0^2 - \omega^2) + i\omega R}$$
$$\beta(\omega, k_t) = \frac{2M}{\mu_0 S_0^2}(\cos(k_x a) + \cos(k_y a)) \qquad (16)$$

Taking into account that the resonators are excited by TE waves, with no electric field component along the z-axis, the wave impedance $\eta_0$ must be taken as:

$$\eta_0 = \frac{\omega \mu_0}{k} \qquad (17)$$

Thus, after substituting (16) and (17) in (13) and (7), the transmission coefficient or transfer function as given in (2) is calculated.

Fig. 2.a shows a typical plot of the transfer function $T(\omega, k_x, k_y=0)$. The parameters used for the calculation are the same as in the experiments reported in the next section. They are given in the caption of the figure. The ring inductance was obtained from the experimental frequency of resonance of the rings, $\omega_0$, through $\omega_0 = (LC)^{-1/2}$, and the mutual inductance was estimated by assuming that each ring is seen by its closest neighbors as a point dipole. The peaks in the plot correspond to the direct excitation of the two MISWs supported by the device. A flat region of $T \sim 1$ can be clearly appreciated between such peaks. From these results it is apparent that, for imaging applications, the operating frequency of the lens must be chosen between the pass-bands of both MISWs, close to the resonant frequency of the rings, $\omega_0$=137.5 MHz. The transfer function $T(\omega,k_x,k_y=0)$ at $\omega$=140 MHz is shown in Fig. 2.b. It can be clearly seen that this transfer function is very close to unity for almost all values of the wavenumber $k_x$. It is worth to mention that, if the distance between the arrays supporting the MISWs increases, both MISWs branches get closer, and the frequency region where the transfer function is flat gets narrower. If the arrays are far enough, the MISW bands overlaps around the resonant frequency of the rings, and the imaging capabilities of the lens disappear. This effect limits the MI lens thickness and, therefore, the distance over which a source can be imaged.

**III. EXPERIMENT**

The model reported in the previous section has been checked by an experiment. For such experiment a MI lens, operating in the radio-frequency range, was fabricated. The lens consisted of two plane arrays of CLORs, placed parallel and separated by a distance $d$=4.5 mm. The periodicity of the arrays is $a$=12.5 mm. Each array has 14x14 resonators that have been fabricated by photoetching planar open metallic loops on a low-loss dielectric substrate. Surface mounted capacitors (nominally $C$=82 pF) have been soldered in the gap of each loop. The external diameter of each loop is 1 cm, and the width of the metallic strips is 1 mm. The experimental setup, including a detail of the ring resonators, is shown in Fig. 1. The frequencies of resonance of the rings on the array were measured, and it was found that all of them fall inside the range $\omega/2\pi$= 137.5 MHz (within a 2\%). The ring self-inductance was calculated from this value, and from the value of the aforementioned capacitance $C$=82 pF. This calculation provided the value $L$=16.3 nH. The source was a square shaped narrow parallel-strips transmission line, short circuited at the end. This source is shown in Fig. 3. The total length of the transmission line is approximately $8a$=10 cm, much smaller than the wavelength at 140 MHz, which is around 200 cm. Therefore, the current can be considered almost uniform along the transmission line. The parallel-wire line was sharp bended in order to produce a strong and uniform square-shaped magnetic field crest in the source plane. Each side length of this square-shaped crest is 2 $a$ ($a$ is the distance between two consecutive rings in the MI lens). The probe used for field measurements was the small loop antenna shown in Fig. 3. The loop radius was 4 mm., and the measured input reactance of such antenna was 10.7 $\Omega$, well below the value of the characteristic impedance of the input and output coaxial waveguides $Z_0$ =50 $\Omega$. This value of the loop reactance ensures that the transmission coefficient between the source and the probe actually corresponds to a measurement of the field [19]. In the experiments, the source is placed in front of the lens (at different distances from it), so that the magnetic crest passes through the centers of 8 consecutive rings, leaving a single ring at its center (see Fig. 3). The image plane is fixed at the exit interface of the lens (as it was already mentioned, this configuration improves the resolution [16]). For the measurements, the probe is consecutively placed at the center of each ring, at the exit interface of the lens. The measured transmission coefficient between the source and the probe provides a map of the field at the image plane. An Agilent Technologies Network Analyzer E8363B was used to measure the transmission coefficient between the source and the probe.

Before to proceed with the direct measurement of any image, the frequency at which such image is formed must be determined. For this purpose, the probe was placed at the center of the ring placed just at the center of the source. A plot of the measured frequency dependence of the transmission coefficient is shown in Fig. 4. We interpret that the two peaks centered at 130 MHz and 150 MHz correspond to the excitation of the two MISWs branches predicted by Fig. 2. The middle point in the curve of Fig. 4 corresponds to a frequency of approximately 140 MHz, which is also the frequency for the theoretical plot in Fig. 2.b. Once the frequency for the image formation was determined, the image formed at the exit interface of the lens, with the source placed at different distances, was measured. The results of such measurements at the frequency of 140 MHz, made following the method described above, are shown in Fig. 5. Fig. 5.a shows a map of the field intensity at the source plane, without the lens. Fig. 5.b shows a map of the field intensity at the exit interface of the lens, when the source is placed at a distance $2d$ from this interface. Finally, Fig. 5.c shows a map of the field at the exit interface of the lens, when the source is placed at a distance of $1.5d$ from such interface. The scale in all Figures is the same and, for data collection, the probe was moved by identical steps (of length *a*) along the co-ordinate axes in the plane of the measurements. The formation of a clear image of the source can be observed in Fig. 5.b, where it can be also seen that the field intensity at the source is approximately reproduced at the image. This results is in complete agreement with the theoretical model. The "image" seen in Fig. 5.c seems to reproduce approximately the form of the source, but with a substantial enhancement (at the cost of some distortion) of the field intensity.

In order to have a better picture of the lens resolution, the field intensity along the line marked in Fig. 3, at different distances, is plotted in Fig. 6. For comparison purposes, two plots of the field intensity along this line at a distance $2d$ from the source, and in the source plane, are also included in the Figure. The curve corresponding to the field intensity measured at the exit interface of the lens, when the source is placed at a distance $2d$ from this interface, approximately matches the curve showing the field intensity at the source plane. This fact corroborates the theoretical predictions about image formation in the MI lens. The curves corresponding to smaller distances between the source and the image, clearly show an enhancement of the image, which is in agreement with some previous theoretical predictions [20], and may be useful for some practical applications. Finally, the overall effect of the lens on the image formation can be appreciated by comparing with the curve showing the field intensity in air, at a distance $2d$ from the source (in the absence of the lens).

It is of interest to investigate the behavior of the lens at those frequencies where the MISWs shown in Fig. 4 can be excited. The "image" formed when the source is placed at a distance $2d$ from the image plane (the exit interface of the lens) is shown in Fig. 7 for the frequency of 130MHz. It can be clearly seen that, in spite of the high transmission level at the central ring, the image is not formed at all. This result clearly shows the importance of avoiding the excitation of MISWs if good images are desired. For this purpose, both arrays must be strongly coupled, so that the frequency band-pass of both MSSWs do not overlap. This fact is the main limitation for manufacturing MI lens with larger thickness, which would allow for image formation at larger distances from the source. For this purpose, it will be desirable to reduce the band-width of the MISWs supported by the lens. Since this band-width is mainly given by the *M/L* ratio [14], a good strategy may be to design magnetic resonators with higher self-inductances, while keeping constant the mutual inductance. The use of swiss-rolls [20] instead of capacitively loaded open rings seems to be a promising alternative in such direction. Future works will explore this and other possibilities.

## IV. CONCLUSION

Image formation in MI lenses has been studied theoretically and experimentally. A theoretical model has been developed, which predicts the main features of image formation in such lenses. This model makes clear the analogy existing between surface plasmons in negative refractive slabs and MISWs waves in MI lenses. A key conclusion of the model is that, for a correct image formation, the MI lens should be thin enough to avoid the overlapping of the two MISWs branches supported by the lens. If this condition is fulfilled, the model shows that the image is formed at frequencies near the resonant frequency of the rings, where no MISWs can be excited in the device. The theoretical model has been validated by experiments showing the image formation at the predicted frequencies. They also showed how the lens losses its imaging properties at those frequencies where MISWs are excited in the device. Such experiments also showed a significant image enhancement when the distance between the source and the image planes is smaller than twice the lens thickness. In our opinion, the reported theoretical and experimental results open the door for the design of practical sub-diffraction imaging devices in the megahertz range of frequencies, which may be of interests for magnetic resonance imaging applications.

**ACKNOWLEDGEMENTS**
This work has been supported by DGI, Ministerio de Educación y Ciencia (SPAIN), under project contract TEC2004-04249-C02-02. Authors want to thank to people who participate in the \emph{1[\textrm st] Intern. Workshop on MI waves} (Nov. 16--19, 2005, Osnabruek, Germany) for helpful discussions that partially motivated this work. Thanks also to our assistant of laboratory, Esperanza Rubio, for her help in the fabrication of the lens used in the experiments.

**CAPTION TO FIGURES**

Figure 1 (a): Schematic of the experimental procedure. (b): Detail of the capacitively loaded rings forming the MI lens.

Figure 2. (a): Three dimensional plot of the transfer function for the lens of the experiment with the source placed at a distance $2d$ is the lens thickness) from the image plane, at the exit of the lens. The transfer function is calculated for the following parameters: $a$=12.5 mm, $R$=0.04 Ω, $C$=82 pF, $L$=16.3 nH, $\omega/2\pi$=137.5 MHz and $M$=-0.015 $L$ (the mutual inductance is negative because the magnetic field originating in a ring must change direction in order to cross another ring). The external radius of the resonator is 5 mm and the width of the ring is 1 mm, so that the effective area of the resonator is $S_0=\pi r^2$, where an averaged radius $r$=4.5 mm is considered. (b): Two-dimensional plot of the transfer function at $\omega/2\pi$=140 MHz.

Figure 3. Photograph of the antennas used as source and probe. A paper sheet with the ring contours drawn on it was placed below the source antenna, in order to illustrate their relative location during the experiments. The dashed line shows the line along which the measurements of Fig. 6 were done.

Figure 4. Plot of the transmission coefficient between the source and the probe, when the probe is located over the central ring of the lens, and the source at a distance $d$ from the exit interface of the lens.

Figure 5. (a): Map of the z-component of the magnetic field intensity at the source plane. (b): Map of the z-component of the magnetic field intensity at the image plane (exit of the lens), when the source is located at a distance $2d$ from the image plane. (c): Map of the z-component of the magnetic field intensity at the image plane (exit of the lens), when the source is located at a distance $1.5d$ from the image plane.

Figure 6. Field profiles along the Y=0 line of Fig. 3 for different distances between the source and the image plane (i.e., at the exit of the lens): $d$ (■), $1.5d$ (▲) and $2d$ (▼). The field at the source plane (black solid line without symbols), as well as the field measured in air (●), without the lens, at a distance $2d$ from the source plane are also shown.

Figure 7. "Image" formed at the exit interface of the lens, under the same circumstances as in Fig. 5.b, but at the frequency $\omega/2\pi=130$ MHz, corresponding the left peak of Fig. 4.

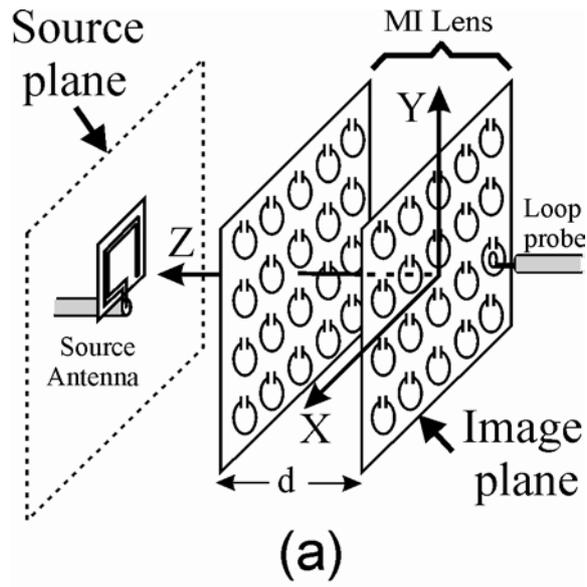

(a)

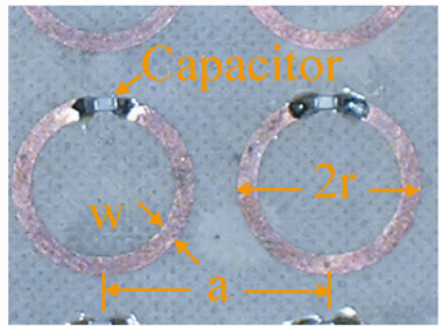

(b)

Figure 1.

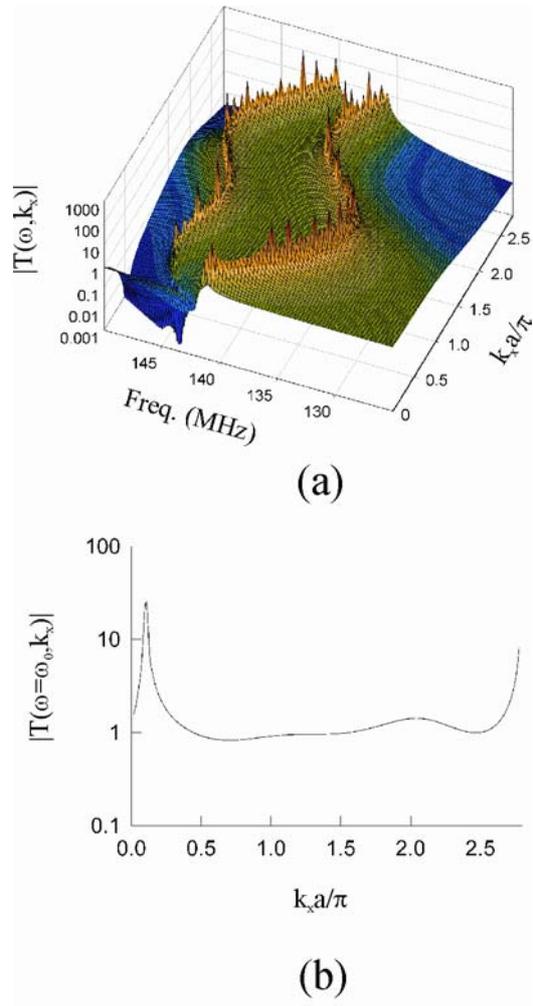

Figure 2

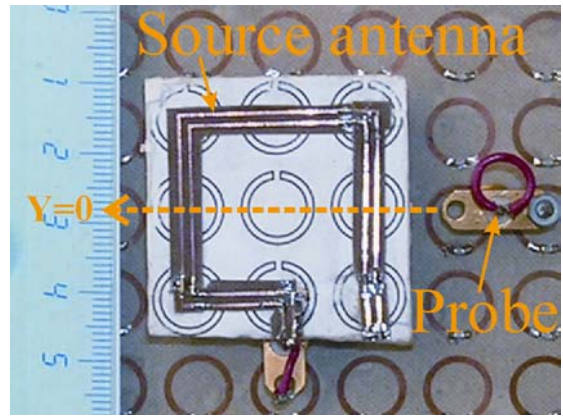

Figure 3

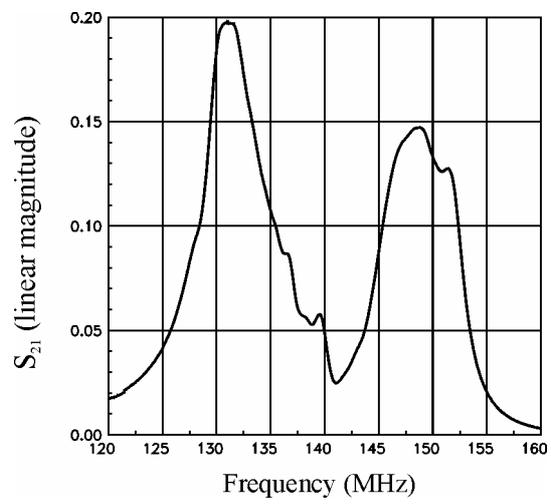

Figure 4

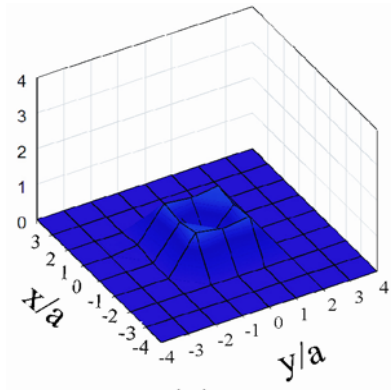

(a)

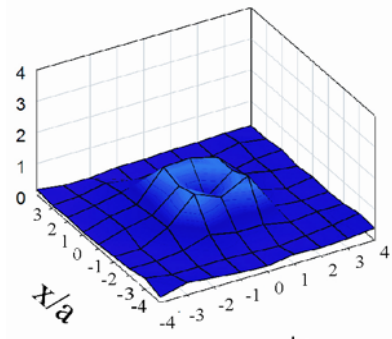

(b)

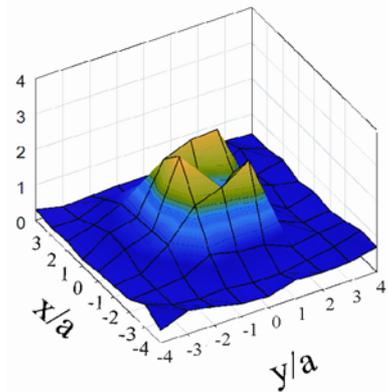

(c)

Figure 5

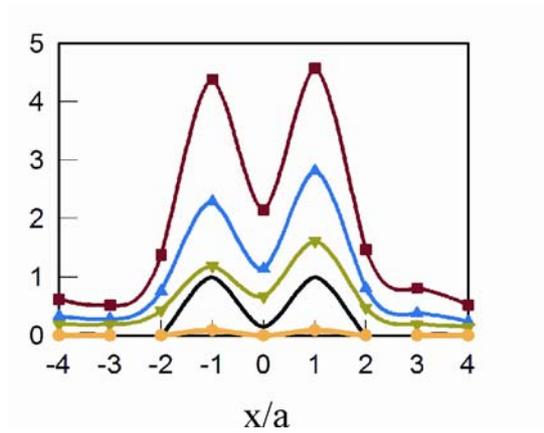

Figure 6

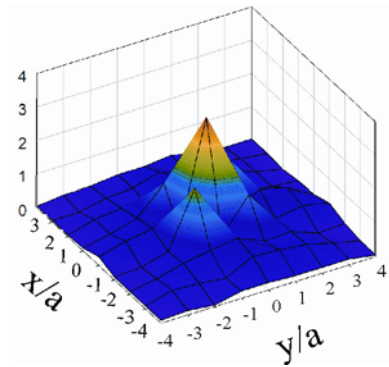

Figure 7